\documentclass[aps,prd,twocolumn,showpacs,preprintnumbers,amsmath,amssymb]{revtex4}

\usepackage{graphicx}
\usepackage{dcolumn}
\usepackage{bm}

\begin{document}

\title{The $\Delta$ Contribution to the Parity-violating Nucleon-nucleon Force
\thanks{Supported by the National Natural Science
Foundation of China under Grants 10625521, 10721063 and 10775146,
the China Postdoctoral Science Foundation (20070420526) and Ministry
of Education of China.}}

\author{Yan-Rui Liu$^1$}
\email{yrliu@ihep.ac.cn}

\author{Shi-Lin Zhu$^2$}
\email{zhusl@phy.pku.edu.cn}

\affiliation{$^1$Institute of High Energy Physics, P.O. Box 918-4,
Beijing 100049, China \\ $^2$Department of Physics, Peking
University, Beijing 100871, China}

\begin{abstract}
Because the nucleon may be excited and transformed into a virtual
$\Delta$ resonance easily, we consider the decuplet contribution to
the parity-violating (PV) nucleon-nucleon interaction in the chiral
effective field theory. The effective PV nucleon-nucleon potential
is derived without introducing any unknown coupling constants.
\end{abstract}

\maketitle

\section{Introduction}\label{sec1}

The strangeness-changing weak processes ($\Delta S=1$) can be
studied both in the semi-leptonic decays and strangeness-changing
hadronic weak interactions. In contrast, the nuclear parity
violation is almost the unique way to study the $\Delta S=0$
hadronic weak interaction experimentally. Up to now, our knowledge
on such processes is still relatively poor.

The search for nuclear parity violation \cite{Tanner} started
shortly after the observation of the parity violation in nuclear
beta decay \cite{CSWu}. Thereafter, there had been many experimental
investigations of nuclear parity-violation such as the polarized
proton-nucleus scattering, radiative $np$ capture, $\gamma$ decay of
nuclei, neutron spin rotation, and atomic parity-violation
\cite{NPVrev,NPVrev2,NPVrev3,NPVrev-n}.

The parity-violating (PV) effect is very small in nuclear processes.
Such an effect can be measured through the asymmetry of the
cross-sections in the polarized experiments. In some heavy nuclei,
there exist two energy levels with different parity which are very
close to each other. The PV weak interaction mixes these two levels.
Then the asymmetry may be amplified.

However, the few-body nuclear system provides a much cleaner place
to study nuclear parity-violation though the asymmetry is only $\sim
10^{-7}$. Experimental progress in this field is very encouraging.
Several years ago, the longitudinal analyzing power of $\vec{p}p$
scattering was measured at TRIUMF \cite{TRIUMF,TRIUMF2}. There are
also on-going experiments to measure the photon asymmetry in
radiative $\vec{n}p$ capture at LANSCE \cite{NPDGamma}, the helicity
asymmetry in the photodisintegration of deuterium at IASA
\cite{IASA}, and the spin rotation of polarized neutrons in $^4$He
at NIST \cite{NIST,NIST2}.

Strong interaction dominates the nucleon-nucleon interaction, which
is repulsive at the short range. Therefore the weak interactions
between nucleons mediated directly by W and Z bosons are strongly
suppressed since the interaction range is around 0.002 fm. On the
other hand, the meson nucleon interaction vertex can be
parity-violating. Thus one can study nuclear parity violation after
replacing one strong vertex by the weak one in the meson exchange
model.

Historically, the study of nuclear parity-violation with the
meson-exchange model started in 1964 \cite{Michel}. Later, nuclear
parity-violation was extensively studied in this framework
\cite{Exmeson,Exmeson2,Exmeson3}. In 1980, Desplanques, Donoghue,
and Holstein (DDH) investigated the PV nuclear force in a general
way and considered the exchanged mesons up to $\rho$ and $\omega$
\cite{DDH}. The DDH method has become a standard way in analyzing
experiments since then. In that paper, the PV vertices were
parameterized with seven coupling constants: $h_\pi^1$,
$h_\rho^{0,1,2}$, $h_\omega^{0,1}$ and $h_\rho^{1\prime}$.
$h_\rho^{1\prime}$ was found to be small and usually neglected
\cite{hrho1p}. DDH estimated these coupling constants using the
quark model and $SU(6)_w$ symmetry. They gave reasonable ranges for
the couplings and presented their best guesses. Surprisingly,
various experimental constraints are more or less consistent with
these DDH estimates, except that the bound on $h_\pi^1$ from nuclear
anapole moment in Cesium \cite{EXAM} does not agree well with those
from other experiments \cite{NAM}.

In the past decades, there has been important progress in the study
of parity violation both in the single-nucleon case
\cite{PVLag,ZhuPHR01,PVSingleN,PVSingleN2,PVSingleN3} and $NN$
system \cite{PVNN,PVNN2,PVNN3,PVNN4,PVNN5}. In order to investigate
nuclear parity violation in a model-independent way, Zhu, Maekawa,
Holstein, Ramsey-Musolf and van Kolck reformulated the PV
nucleon-nucleon interaction in the framework of effective field
theory (EFT) \cite{NPVEFT}. At very low energy, the momenta of
external fields are very small and the pion can be integrated out.
EFT without explicit pions is appropriate. When the external momenta
are comparable with the pion mass, EFT with explicit pions is
necessary.

For the description of PV NN forces in EFT with explicit pions, the
treatment is similar to the study of parity-conserving (PC) NN force
in EFT \cite{EFTforce,EFTforce2}. One simply replaces one PC vertex
with one PV vertex and imposes chiral symmetry on the PV vertex. In
Ref. \cite{NPVEFT}, the PV potential was calculated to ${\cal O}(Q)$
in Weinberg's power counting where $Q$ is the typical scale of the
processes. The leading order (${\cal O}(Q^{-1})$) result reproduced
the pion exchange part of DDH formalism. At the next leading order
(${\cal O}(Q^0)$), explicit computation shows there is in fact no
contribution. At the third order, the short range potential was
described with contact interactions. The medium range potential was
deduced from two-pion exchange interactions while the long range
potential was obtained by considering corrections to the one-pion
exchange interaction. In this framework, Ref. \cite{CPLiu06} studied
a minimal set of parameters to describe low-energy PV observables.
In Ref. \cite{TPEPV}, the authors studied PV asymmetry in $np\to
d\gamma$ within EFT.

The decuplet baryon $\Delta$ couples to $N\pi$ strongly. The virtual
$\Delta$ may aslo contribute to the PV nucleon-nucleon interactions,
which was noted long time ago in Refs.
\cite{roleDelta,roleDelta2,roleDelta3}. The DDH formalism was
extended to investigate the effects due to $\Delta$ \cite{DeltaPVN}.
With development of the modern EFT language, we will extend the
former work \cite{NPVEFT} and calculate the PV potential by
considering $\Delta$ as an explicit degree of freedom in EFT in the
present work. The present work was part of Y. R. Liu's thesis
submitted in April, 2007. It is interesting to note that an
independent work dealing with similar topics appeared recently
\cite{NPVK}. However the way to derive the potential in this work is
different from that in Ref. \cite{NPVK}. In a recent work
\cite{PNCH}, the calculation of the longitudinal asymmetry in $pp\to
pp$ by including $2\pi$ exchange effects which include $NN$ and
$N\Delta$ intermediate states is presented.

In order to include the $\Delta$ degree of freedom systematically,
we employ the heavy baryon chiral perturbation theory with $\Delta$.
The expansion scheme was called the small scale expansion (SSE)
\cite{SSE}, which was widely used to study the processes involving
$\Delta$
\cite{SSEapp,SSEapp2,SSEapp3,SSEapp4,SSEapp5,SSEapp6,SSEapp7,SSEapp8,SSEapp9}.
Both the pion mass and the mass difference between nucleon and
$\Delta$ isobar are counted as the order ${\cal O}(Q)$. We use this
formalism to calculate the $\Delta$ contribution to the PV NN
potential. In the following section, we present the relevant
Lagrangian. In Section {\ref{sec3}}, we calculate PV potentials due
to the virtual $\Delta$ baryon. The final section is a short
summary.
\\

\section{Lagrangians}\label{sec2}

In the EFT study of the nucleon-nucleon potential, one performs a
systematic expansion of Lagrangians and amplitudes
\cite{EFTforce,EFTforce2}. We present relevant Lagrangians ${\cal
L}^{(\nu)}$ in this section. They are grouped with chiral index
$\nu=d+f/2-2$ where $d$ is the number of derivatives and powers of
the pion mass and $f$ the number of fermion fields. When we consider
the $\Delta$ contribution to the parity-violating potential up to
the third order ${\cal O}(Q)$, we need only the lowest order chiral
Lagrangians.

For the $\pi NN$ interaction, the PC part is
\begin{eqnarray}
{\cal L}^{(0)}_{\pi N,PC}&=& \overline{N}[iv \cdot {\cal D} + 2g^0_A
S \cdot A]N,
\end{eqnarray}
where
\begin{eqnarray}
&{\cal D}_\mu = D_\mu + V_\mu, \qquad V_\mu=\frac12(\xi D_\mu
\xi^\dagger +\xi^\dagger D_\mu \xi),&\nonumber\\
&A_\mu=-\frac{i}{2}(\xi D_\mu \xi^\dagger -\xi^\dagger D_\mu
\xi)=-\frac{D_\mu\pi}{F_\pi}+{\cal O}(\pi^3),&\nonumber\\
&\xi=\exp(\frac{i\pi^a \tau^a}{2F_\pi})=\exp(\frac{i\pi}{F_\pi}).&
\end{eqnarray}
Here $V_\mu$ and $A_\mu$ are the chiral connection and the axial
field respectively. $v_\mu$ is the velocity and $S_\mu$ is the
Pauli-Lubanski spin vector. $F_\pi=92.4$ MeV is the pion decay
constant and $\tau^a$ is the Pauli matrix. Here $g_A\simeq 1.27$ is
the nucleon axial vector coupling constant.

The PV part is
\begin{eqnarray}
{\cal L}^{(-1)}_{\pi N,PV}&=& -\frac{h^1_\pi F_\pi}{2\sqrt2}\overline{N}X^3_- N \nonumber\\
&=&-i h^1_\pi(\overline{p}n\pi^+ -\overline{n}p\pi^-)+... ,
\end{eqnarray}
where
\begin{equation*}
X^3_-=\xi^+ \tau^3\xi-\xi \tau^3\xi^+ ,
\end{equation*}
and $h^1_\pi\sim 10^{-7}$ is the weak coupling constant. The
ellipsis denotes terms involving more pions.

For the part containing $\Delta$, the leading Lagrangian reads
\cite{SSE}
\begin{eqnarray}
{\cal L}^{(0)}_{\pi N\Delta,PC}&=& -i \overline{T}^{\mu a}v\cdot
D^{ab}T_\mu^b + \delta \overline{T}^{\mu a}T_\mu^a \nonumber\\
&&+ 2 g_{\pi N\Delta}(\overline{T}^{\mu a}A_\mu^a N +
\overline{N}A_\mu^a T^{\mu a} ),
\end{eqnarray}
where $\delta=m_\Delta-m_N$, $A_\mu^a=\frac12{\rm Tr}(A_\mu \tau^a)$
and $T^\mu$ represents $\Delta$ fields with
\begin{eqnarray}
&T_\mu^1= \frac{1}{\sqrt2}
\left(\begin{array}{c}\Delta^{++}-\Delta^0/\sqrt3\\
\Delta^+/\sqrt3-\Delta^-\end{array} \right)_\mu,&\nonumber\\
&T_\mu^2= \frac{i}{\sqrt2}
\left(\begin{array}{c}\Delta^{++}+\Delta^0/\sqrt3\\
\Delta^+/\sqrt3+\Delta^-\end{array} \right)_\mu,&\nonumber\\
&T_\mu^3= -\sqrt{\frac23}
\left(\begin{array}{c}\Delta^{+}\\
\Delta^0\end{array} \right)_\mu.&
\end{eqnarray}
In this Lagrangian, we have used ${\cal C}=\sqrt2 g_{\pi N\Delta}$
with the language in Ref. \cite{HBChPT,HBChPT2}. The quark model
gives the relation $g_{\pi N\Delta}=\frac{3\sqrt2}{5}g_A$. Since the
PV $\pi N\Delta$ part contributes to the PV potential beyond the
order of ${\cal O}(Q)$ \cite{ZhuPHR01}, we do not consider it here.

\section{$\Delta$ contribution to PV NN potential}\label{sec3}

\begin{figure}
\begin{center}
\scalebox{0.4}{\includegraphics{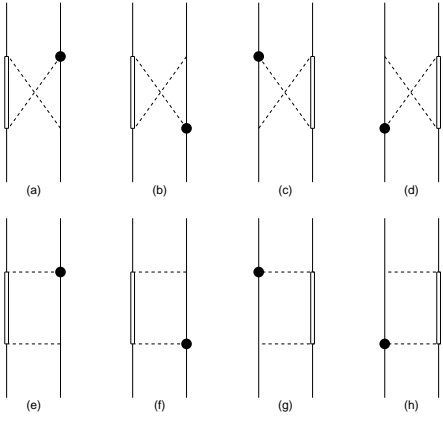}} \caption{Diagrams for
intermediate $\Delta$ contribution to PV NN potential. The dotted
lines are pions. The full lines represent nucleon while the double
lines represent $\Delta$ states. Vertices with black dot mean the
parity-violating $\pi N$ interaction.}\label{fynpvd}
\end{center}
\end{figure}

Because the PV contribution is tiny, one PV vertex is enough for the
present study. The intermediate $\Delta$ contribution to
parity-violating potential is presented in Fig. \ref{fynpvd}. We
employ the counting scheme of SSE and truncate the expansion at the
order ${\cal O}(Q)$. To this order, the triangle diagrams do not
contribute. In the case without the $\Delta$ contribution, the box
diagrams are two-particle reducible. Now the diagrams are all
two-particle irreducible (2PI). That is, the diagrams in Fig.
\ref{fynpvd} will not induce double counting problem. In the
following, we calculate the effective potentials in detail.

First, we consider the cross diagrams (a)-(d) in Fig. \ref{fynpvd}.
From the vertices, one can construct four cases of transitions which
include charge-conserving cases $pp\to pp$, $nn\to nn$ and $pn \to
pn$ ($np\to np$) and charge-changing case $pn\to np$ ($np\to pn$).

For $pp\to pp$, the sum of (a) and (b) gives
\begin{eqnarray}
i T=-i\frac{4\sqrt2}{9}\frac{g_{\pi
N\Delta}^2g_Ah^1_\pi}{\Lambda_\chi^2 F_\pi}Z(Q) \bar{p}[S_1\cdot
q,S_{1\,\mu}]p\bar{p} S_2^\mu p\,.
\end{eqnarray}
where $q=p_1-p_1^\prime=p_2^\prime-p_2$,
$Q^2=-q^2\approx\mathbf{q}^2$, $\Lambda_\chi=4\pi F_\pi$ and
\begin{eqnarray}
Z(Q)&=& 2
L(Q)+\frac{\pi}{2\delta}(4m_\pi^2+Q^2)A(Q)-\frac{2}{\delta}B(Q)\,,
\end{eqnarray}
with
\begin{eqnarray}
L(Q)&=&\frac{\sqrt{4m_\pi^2+Q^2}}{Q}\ln\frac{Q+\sqrt{4m_\pi^2+Q^2}}{2m_\pi}
\,\nonumber\\
A(Q)&=&\frac{1}{2Q}\arctan\frac{Q}{2m_\pi}\,\nonumber\\
B(Q)&=&\int_0^1d\,y\int_\delta^\infty d\lambda
\frac{m_\pi^2-\delta^2+y(1-y)Q^2}{\lambda^2+m_\pi^2-\delta^2+y(1-y)Q^2}\,..
\end{eqnarray}
In calculating the loop integrals, we have used the dimensional
regularization. The divergent part could be absorbed by the
renormalization of the counter-terms at the same chiral order. Here
we only retain the non-analytic terms.

For $nn\to nn$ channel, the sum of (a) and (b) is
\begin{eqnarray}
i T=i\frac{4\sqrt2}{9}\frac{g_{\pi
N\Delta}^2g_Ah^1_\pi}{\Lambda_\chi^2 F_\pi}Z(Q) \bar{n}[S_1\cdot
q,S_{1\,\mu}]n\bar{n} S_2^\mu n\,.
\end{eqnarray}
This result is similar to the $pp\to pp$ channel. Similarly, one
gets contributions from the mirror diagrams (c) and (d). Since the
initial particles are identical, the operator form for these two
channels will generate (a), (b) and the mirror diagrams (c), (d)
simultaneously.

Compared with the case without the $\Delta$ contribution
\cite{NPVEFT}, there is an additional channel $pn\to pn$. The sum of
(a)-(d) in the operator form reads
\begin{eqnarray}
i T&=&i\frac{2\sqrt2}{3}\frac{g_{\pi
N\Delta}^2g_Ah^1_\pi}{\Lambda_\chi^2 F_\pi}Z(Q)\Big\{ \bar{N}[S\cdot
k,S_\mu]\tau_3 N\bar{N} S^\mu N\nonumber\\
&&-\bar{N}[S\cdot k,S_\mu]N\bar{N} S^\mu \tau_3 N\Big\}\,.
\end{eqnarray}
Here $k$ is the initial momentum minus the final momentum for a
nucleon line.

After combining the above three channels, we get
\begin{eqnarray}\label{crossconserveflavor}
i\frac{4\sqrt2}{9}\frac{g_{\pi N\Delta}^2g_Ah^1_\pi}{\Lambda_\chi^2
F_\pi}Z(Q) \bar{N}[S\cdot
k,S_\mu]\tau_3N\bar{N} S^\mu N\nonumber\\
-i\frac{8\sqrt2}{9}\frac{g_{\pi N\Delta}^2g_Ah^1_\pi}{\Lambda_\chi^2
F_\pi}Z(Q) \bar{N}[S\cdot k,S_\mu]N\bar{N} S^\mu \tau_3 N\,.
\end{eqnarray}

For the charge-changing case $pn\to np$, the sum of diagrams (a)-(d)
gives
\begin{eqnarray}\label{crosschangeflavor}
-\frac{\sqrt2}{6}\frac{g_{\pi N\Delta}^2 g_A
h^1_\pi}{\Lambda_\chi^2F_\pi} Y(Q)\epsilon^{ij3}\bar{N}\tau^i
N\bar{N} \tau^j \boldsymbol{\sigma}\cdot\mathbf{k} N \,,
\end{eqnarray}
where
\begin{eqnarray}
Y(Q)&=&2L(Q)+\frac{2\pi}{3\delta}(2m_\pi^2+Q^2)A(Q)-\frac{2}{\delta}C(Q)
\,,
\end{eqnarray}
with
\begin{eqnarray}
C(Q)&=&\int_0^1d\,y\int_\delta^\infty d\lambda
\frac{m_\pi^2-\delta^2+\frac43y(1-y)Q^2}{\lambda^2+m_\pi^2-\delta^2+y(1-y)Q^2}\,.
\end{eqnarray}

Next, we consider the box diagrams (e)-(h) in Fig. \ref{fynpvd}.
There are also four cases: charge-conserving processes $np\to np$
($pn\to pn$), $pp\to pp$ and $nn\to nn$ and charge-changing process
$np\to pn$ ($pn\to np$).

For the channel $np\to np$, the sum of the diagrams (e)-(h) gives
\begin{eqnarray}
i T&=&i\frac{2\sqrt2}{9}\frac{g_{\pi
N\Delta}^2g_Ah^1_\pi}{\Lambda_\chi^2 F_\pi}W(Q)\Big\{ \bar{N}[S\cdot
k,S_\mu]\tau_3N\bar{N} S^\mu N\nonumber\\
&&-\bar{N}[S\cdot k,S_\mu]N\bar{N} S^\mu \tau_3N\Big\}\,,
\end{eqnarray}
where
\begin{eqnarray}
W(Q)=2L(Q)-\frac{\pi}{2\delta}(4m_\pi^2+Q^2)A(Q)-\frac{2}{\delta}B(Q)\,.
\end{eqnarray}
In calculating the amplitudes, we use the following formula
\begin{eqnarray}
\frac{1}{v\cdot k+i\epsilon}=-\frac{1}{-v\cdot k+i\epsilon}-2\pi
i\delta(v\cdot k)\,.
\end{eqnarray}
In the case without the $\Delta$ contribution, the part from the
$\delta$ function was subtracted to separate the contributions from
the iterated one-pion exchange and those from the irreducible
two-pion exchange. Now this part is included because the diagram is
2PI.

The diagrams (e) and (f) result in
\begin{eqnarray}
i T &=& -i\frac{4\sqrt2}{3}\frac{g_{\pi
N\Delta}^2g_Ah^1_\pi}{\Lambda_\chi^2 F_\pi}W(Q) \bar{p}[S_1\cdot
q,S_{1\,\mu}]p\bar{p} S_2^\mu p \,,
\end{eqnarray}
for the channel $pp\to pp$ and
\begin{eqnarray}
i T &=& i\frac{4\sqrt2}{3}\frac{g_{\pi
N\Delta}^2g_Ah^1_\pi}{\Lambda_\chi^2 F_\pi}W(Q) \bar{n}[S_1\cdot
q,S_{1\,\mu}]n\bar{n} S_2^\mu n
\end{eqnarray}
for $nn\to nn$.

After combining these results, we get the charge-conserving
amplitude from box diagrams
\begin{eqnarray}\label{boxconserveflavor}
-i\frac{4\sqrt2}{9}\frac{g_{\pi N\Delta}^2g_Ah^1_\pi}{\Lambda_\chi^2
F_\pi}W(Q) \bar{N}[S\cdot
k,S_\mu]\tau_3N\bar{N} S^\mu N\nonumber\\
-i\frac{8\sqrt2}{9}\frac{g_{\pi N\Delta}^2g_Ah^1_\pi}{\Lambda_\chi^2
F_\pi}W(Q) \bar{N}[S\cdot k,S_\mu]N\bar{N} S^\mu \tau_3 N\,.
\end{eqnarray}

For the charge-changing case $np\to pn$, one sums the amplitudes
from (e)-(h) and gets
\begin{eqnarray}\label{boxchangeflavor}
-\frac{\sqrt2}{6}\frac{g_{\pi N\Delta}^2 g_A
h^1_\pi}{\Lambda_\chi^2F_\pi} X(Q)\epsilon^{ij3}\bar{N}\tau^i
N\bar{N} \tau^j \boldsymbol{\sigma}\cdot\mathbf{k} N \,,
\end{eqnarray}
where
\begin{eqnarray}
X(Q)=2L(Q)-\frac{2\pi}{3\delta}(2m_\pi^2+Q^2)A(Q)-\frac{2}{\delta}C(Q)
\,.
\end{eqnarray}

After combing Eqs. (\ref{crossconserveflavor}),
(\ref{crosschangeflavor}), (\ref{boxconserveflavor}) and
(\ref{boxchangeflavor}), we finally get the $\Delta$ contribution to
nuclear parity violation
\begin{eqnarray}\label{totalcorrection}
-\frac{\sqrt2}{9}\frac{g_{\pi N\Delta}^2g_Ah^1_\pi}{\Lambda_\chi^2
F_\pi}\Big[W(Q)-Z(Q)\Big]\epsilon^{ijk} N^\dagger k^i\sigma^j\tau_3NN^\dagger\sigma^k N\nonumber\\
-\frac{2\sqrt2}{9}\frac{g_{\pi N\Delta}^2g_Ah^1_\pi}{\Lambda_\chi^2
F_\pi}\Big[W(Q)+Z(Q)\Big]\epsilon^{ijk} N^\dagger k^i\sigma^jNN^\dagger \sigma^k \tau_3 N\nonumber\\
-\frac{\sqrt2}{6}\frac{g_{\pi N\Delta}^2 g_A
h^1_\pi}{\Lambda_\chi^2F_\pi}
\Big[X(Q)+Y(Q)\Big]\epsilon^{ij3}N^\dagger\tau^i NN^\dagger \tau^j
\boldsymbol{\sigma}\cdot\mathbf{k} N \,.
\end{eqnarray}
Acordingly, one gets the PV potential
\begin{eqnarray}
V&=&-\frac{i}{\Lambda_\chi^3}\left\{\tilde{C}_2^\Delta(Q)
\frac{\tau_1^z+\tau_2^z}{2}(\boldsymbol{\sigma_1\times\sigma_2})\cdot\mathbf{q}\right.\nonumber\\
&&\left.+ C_6^\Delta(Q)(\boldsymbol{\tau_1\times\tau_2})^z
(\boldsymbol{\sigma_1+\sigma_2})\cdot\mathbf{q}\right\} \,,
\end{eqnarray}
where
\begin{eqnarray}
\tilde{C}_2^\Delta(Q) 
&=&\frac{8\sqrt2}{9}\pi g_{\pi N\Delta}^2g_Ah^1_\pi
\Big[4L(Q)-\frac{4}{\delta}C(Q)\Big], \nonumber\\
C_6^\Delta(Q)
&=&-\frac{2\sqrt2}{3}\pi g_{\pi N\Delta}^2g_Ah^1_\pi
\Big[8L(Q)\nonumber\\
&&-\frac{\pi}{\delta}(4m_\pi^2+Q^2)A(Q)-\frac{8}{\delta}B(Q)\Big].
\end{eqnarray}

\section{Discussions}\label{sec4}

\begin{figure}
\begin{center}
\scalebox{0.8}{\includegraphics{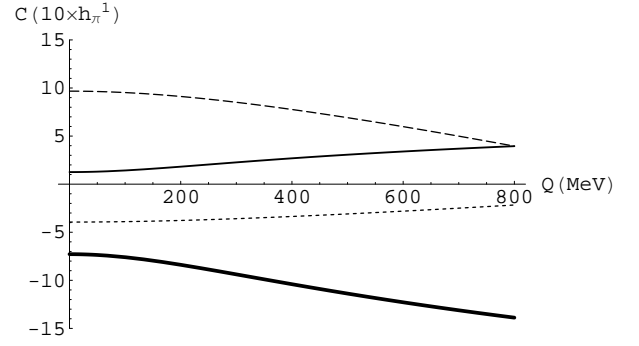}} \caption{The momentum
dependence of coefficients in the PV two-pion exchange potentials:
$\tilde{C}_2^{2\pi}(Q)$ (thick solid line), $C_6^{2\pi}(Q)$ (thin
solid line), $\tilde{C}_2^\Delta(Q)$ (dash line) and $C_6^\Delta(Q)$
(dotted line). }\label{comc}
\end{center}
\end{figure}

In short summary, we have calculated the $\Delta$ contribution to
the parity-violating nucleon-nucleon potential. The $\Delta$
resonance couples to the nucleon and pion strongly. A nucleon may be
excited and transformed into a virtual $\Delta$ quite easily. The
$\Delta$ will certainly contribute to the hadronic parity violation
in nuclear processes. In this work, we employ the small scale
expansion formalism and extend the former investigation of PV NN
interaction in EFT in Ref. \cite{NPVEFT} through the inclusion of
the $\Delta$ contribution. To the next-next-leading order, the new
potential contains no more unknown PV coupling constants. The only
new parameter is the strong coupling constant $g_{\pi N\Delta}$,
which is known from the decay width of the $\Delta$ baryon.

The structure of the obtained potential is similar to the
medium-range potential derived in Ref. \cite{NPVEFT,NPVrev-n}. For
comparison, we plot the momentum dependence of coefficients
$\tilde{C}_2^\Delta(Q)$, $C_6^\Delta(Q)$,
$\tilde{C}_2^{2\pi}(Q)=-8\sqrt2\pi g_A^3h_\pi^1L(Q)$ and
$C_6^{2\pi}(Q)=-\sqrt2\pi
g_Ah_\pi^1L(Q)+\sqrt2\pi[3L(Q)-H(Q)]g_A^3h_\pi^1$ with
$H(Q)=\frac{4m_\pi^2}{4m_\pi^2+Q^2}L(Q)$ in Fig. \ref{comc}. We take
$m_\pi=135$ MeV, $\delta=294$ MeV, $g_A=1.27$, $g_{\pi
N\Delta}=\frac{3\sqrt2}{5}g_A$. From the figure, one notes
$C_6^\Delta(Q)$ is bigger than $C_6^{2\pi}(Q)$ at small momentum. It
is also important to note that $\tilde{C}_2^\Delta(Q)$ and
$\tilde{C}_2^{2\pi}(Q)$ are comparable in magnitude but they have
opposite signs! Therefore it is highly desirable to include the new
parity violating nucleon-nucleon arising from the $\Delta$
correction in the future theoretical calculation of PV observables
in the hadronic weak interaction processes.

\vfill
\section*{Acknowledgments}

We thank Prof. B. Desplanques for helpful discussions. This project
was supported by the National Natural Science Foundation of China
under Grants 10625521, 10721063 and 10775146, the China Postdoctoral
Science Foundation (20070420526) and Ministry of Education of China.

\end{document}